\documentclass[12pt]{article}
\usepackage{amsmath,amssymb,epsfig,cite}
\usepackage{color}

\makeatletter

\@addtoreset{equation}{section} \makeatother

\begin{document}

\titlepage
\begin{flushright}
\end{flushright}
\vskip 1cm
\begin{center}
{\large \bf Brane Cosmology in an Arbitrary Number of Dimensions}
\end{center}

\vspace*{5mm} \noindent

\centerline{N.~Chatillon\footnote{nchatill@physics.syr.edu},
C.~Macesanu\footnote{cmacesan@physics.syr.edu} and
M.~Trodden\footnote{trodden@physics.syr.edu}}

\centerline{ \em Department of Physics, Syracuse University,
Syracuse, NY 13244-1130, USA.} \vskip 0.5cm

\begin{center}
{\bf Abstract}
\end{center}

We derive the effective cosmological equations for a
non-$\mathbb{Z}_2$ symmetric codimension one brane embedded in an
arbitrary D-dimensional bulk spacetime, generalizing the $D=5,6$
cases much studied previously. As a particular case, this may be
considered as a regularized codimension (D-4) brane avoiding the
problem of curvature divergence on the brane. We apply our results
to the case of spherical symmetry around the brane and to partly
compactified AdS-Schwarzschild bulks.

\vskip 1cm

\section{Introduction}
The possibility of extra spatial dimensions, in addition to the
three that we perceive in everyday life, is a crucial aspect of a
number of constructions in modern particle physics, from string
theory to methods to address the hierarchy
problem~\cite{Kaluza:tu,Klein:tv,Rubakov:1983bb,Antoniadis:1990ew,Lykken:1996fj,
Arkani-Hamed:1998rs,Antoniadis:1998ig,Randall:1999ee,Randall:1999vf,Lykken:1999nb,
Arkani-Hamed:1999hk,Antoniadis:1993jp,Dienes:1998vg,Kaloper:2000jb}.
In many of these models, the matter content of our
three-dimensional universe is confined to a submanifold, or
3-brane, while gravity, via the equivalence principle, propagates
in the entire space-time, or bulk.

Large extra dimension models with matter localized on a 4d brane
have been re-popularized recently by their application to the
electroweak hierarchy problem
\cite{Arkani-Hamed:1998rs,Antoniadis:1998ig}. In these models the
bulk metric, typically a flat hypertorus, is assumed to be
unaffected by the energy-momentum content of the brane, which
allows for a formulation independently of the total dimension. At
the background level, this is justified by the absence of a brane
tension. The effect of the brane matter is less trivial in
general. For energies small compared to the typical inverse radius
$R^{-1}$ of the bulk, the Kaluza-Klein (KK) zero-mode truncation
applies as a valid approximation for the bulk metric. The
non-trivial curvature induced on the bulk metric may be studied
above these energies by including the tower of non-zero KK
gravitons as effective 4d particles. Brane-world cosmology with
flat bulk has received much careful study, although previous work
has mostly focused on the low-energy regime (although far above
$R^{-1}$), under a ``normalcy temperature" $T_*$ evaluated in
\cite{Arkani-Hamed:1998nn, Macesanu:2004gf}, where the bulk size
is assumed to have been already stabilized by an unspecified
high-energy mechanism and the non-zero KK gravitons do not have a
significant effect on the evolution of the brane metric. The
effect of brane matter on the bulk metric evolution at earlier
times, and the consequent alteration of the brane metric
evolution, has not yet been studied systematically in an arbitrary
number of dimensions.

The problem extends to unwarped brane models with non-flat but
homogeneous transverse space. The conclusions of cosmological
studies where the brane is a point in a compact hyperbolic
transverse space \cite{Starkman:2001xu,Starkman:2000dy}, for
instance, are valid only as long as the brane content does not
significantly alter the bulk homogeneity condition.

Another very interesting class of brane models considers the
cosmological evolution resulting from an induced kinetic term for
gravity localized on the brane in addition to the bulk one
\cite{Deffayet:2001xs,Deffayet:2001pu,Dvali:2000hr,
Deffayet:2002sp,Deffayet:2000uy,Dvali:2003rk,Lue:2004rj,Lue:2002sw}
; we focus here on the more simple case of models with pure bulk
gravity.

Compact and non-compact models with extra dimensions warped by the
presence of a large brane tension provide an elementary example
in which the backreaction effect is taken into account. In five
dimensions, the brane is codimension one and leads to an
exponential warping useful for explaining the Standard Model
hierarchies as a gravitational effect \cite{Randall:1999ee}, or for
localizing gravity with non-compact bulks \cite{Randall:1999vf}.
However, here again it is a non-trivial task to treat the backreaction
of a more general cosmological brane source beyond the zero-mode
approximation.

In five and six dimensions, exact and perturbative results have
been obtained for such a source. In the simplest case, the
cosmological evolution of a brane with tension $\lambda$ in an
infinite $D=5$ bulk leads to exactly calculable
$O(\rho^2/\lambda^2)$ corrections to the Friedmann equation
\cite{Binetruy:1999hy}. The compact case is more subtle as one
must also allow for the evolution of the radion
field~\cite{Binetruy:2001tc}. The aberrant Hubble rate, quadratic
in $\rho$ for $D=5$ flat compact bulks
\cite{Binetruy:1999ut,Chung:1999zs} for instance, has been shown
\cite{Csaki:1999mp} to result from the unnatural requirement of a
static radion without introducing a stabilization mechanism, which
leads to an overconstraint on the extra matter on a second brane.
Stabilized models treated perturbatively in $\rho$ again lead to a
correct Friedmann equation, linear in $\rho$ with quadratic
corrections \cite{Cline:2002ht}. We leave the review and
discussion of more general five-dimensional configurations,
including non-$\mathbb{Z}_2$ symmetric bulks and ``dark radiation"
terms, for section \ref{bulkspher}.

The problem is qualitatively different for a codimension two brane
in $D=6$, a configuration reminiscent of the much studied cosmic
string solutions in four dimensions. Indeed, for any brane source
differing from a pure tension, the curvature diverges on the brane
on which observers are supposed to live. This problem has been solved
by considering regularized or ``fat" branes with finite width. In
the simple case of the compact ``football" shaped bulk solution
whose size is stabilized by a bulk magnetic flux
\cite{Carroll:2003db,Navarro:2003vw}, the Friedmann equation
once again exhibits perturbative $\rho^2$ corrections at high energies
\cite{Vinet:2004bk,Vinet:2005dg}. Backreaction effects in
cosmological brane models for $D>6$ remain to be investigated.

Since it is expected that higher codimension infinitely thin
branes will be singular, regularization is necessary, by an
homogeneous spherical core for example. Alternatively, if the
brane matter is taken to be localized at the boundary of this core
(see figure \ref{figure with non-symmetric bulk}), the codimension
(D-4) brane is regularized as an infinitely thin codimension one
brane shell with a (D-1)-dimensional worldvolume. The extra (D-5)
spatial dimensions on the worldvolume in addition to the three
observable ones may then be eliminated by compactification with a
sufficiently small radius\footnote{After completion of this work,
reference \cite{Cuadros-Melgar:2005ex} was pointed to our
attention, following a similar approach for the particular $D=6$
case.}.

In this paper, we derive the effective cosmological equations for
a non-$\mathbb{Z}_2$ symmetric codimension one brane embedded in an
arbitrary D-dimensional bulk spacetime. Our results generalize the
$D=5,6$ cases which have been exhaustively studied previously by other
authors, exactly in the case of infinitely thin branes for $D=5$,
and perturbatively with regularized core branes for $D=6$. As we
shall see, in the case of higher codimension, there is
significantly more choice for the bulk geometry and hence, in the
equivalent 4-dimensional effective theory, for the nature of the
extra terms induced in the Friedmann equation.

The paper is organized as follows. In the next section we review
the derivation of the Einstein equations induced on the brane in
an arbitrary number of dimensions. In section~\ref{bulkspher} we
consider the case of a general unspecified bulk with spherical
symmetry around the brane, including non-$\mathbb{Z}_2$ symmetric
contributions. We then derive a non-conservation law for the dark radiation specific to
$D>5$ brane models, and we apply our
findings to a specific non-compact orbifold bulk. In
section~\ref{adsbulk} we take the example of an AdS-Schwarzschild
bulk in which spherical symmetry is broken by a toroidal
compactification of the transverse space on the brane. In
section~\ref{conclusions} we provide a discussion of our results
and conclude.

\section{Brane gravity in an arbitrary number of dimensions}

\subsection{The Effective Einstein equations}

The brane is taken to be a $(D-1)$-dimensional hypersurface, a
particular member of a foliation of an unspecified $D-$dimensional
spacetime, referred to as {\it the bulk} in what follows. Defining as $n^a$ the
unit spacelike vector normal to the brane\footnote{Our metric
signature is mostly positive, so that $n^2=1$ .}, the induced
metric on the brane is then $\tilde{g}_{ab} = g_{ab} - n_a n_b$.
The Gauss-Codazzi equations, with no
dynamical assumptions, geometrically relate the bulk Riemann tensor ${\cal R}^a_{\
bcd}$ to the brane one $\tilde{\cal R}^m_{\ npq}$ (built from its
induced metric $\tilde{g}_{ab}$) and to the extrinsic
curvature tensor on the brane, given by

\begin{equation}
K_{ab} = K_{ba} = -\tilde{g}^c_{\ a}\ \tilde{g}^d_{\ b}\ \nabla_c n_d \  .
\end{equation}
They read respectively

\begin{eqnarray}
{\cal R}^a_{\ bcd}\ \tilde{g}_a^{\ m} \ \tilde{g}^b_{\ n} \ \tilde{g}^c_{\ p}
\ \tilde{g}^d_{\ q} &=& \tilde{\cal R}^m_{\ npq} + K^m_{\ q} K_{np} -
K^m_{\ p} K_{nq}
\label{Gauss equation}
\\
{\cal R}_{ab}\ \tilde{g}^a_{\ c} \ n^b &=& \tilde{\nabla}_e K^e_{\ c} -
\tilde{\nabla}_c K \ ,
\label{Codazzi equation}
\end{eqnarray}
where $\tilde{\nabla}$ is the covariant derivative adapted to the
brane induced metric.

Including gravitational dynamics, the Gauss equation~(\ref{Gauss
equation}) can be used to covariantly obtain the projected
Einstein equations for the induced metric, to which brane
observers are assumed to be minimally coupled. To achieve this,
one first decomposes the bulk Riemann tensor into its
algebraically irreducible components:

\begin{equation}
{\cal R}^{ab}_{\ \ cd}= C^{ab}_{\ \ cd} +
\frac{1}{D-2}\delta^{[a}_{\ [c} \Big({\cal R}^{b]}_{\
d]}-\frac{1}{D}\delta^{b]}_{\ d]}{\cal R}\Big)
+\frac{1}{2D(D-1)}\delta^{[a}_{\ [c} \delta^{b]}_{\ d]} {\cal R} \
,
\end{equation}
where $M_{[ab]}\equiv M_{ab}-M_{ba}$, and $C^{ab}_{\ \ cd}$ is the
Weyl tensor, possessing the same symmetries as the Riemann tensor,
plus total tracelessness. The Ricci tensor ${\cal R}^a_{\ b}$ is
algebraically related to the bulk energy-momentum tensor ${\bf
T}^a_{\ b}$ through the bulk Einstein equations

\begin{equation}
{\cal R}^a_{\ b} = 8 \pi G_D \Big( {\bf T}^a_{\ b} - \frac{1}{D-2} {\bf T}
\delta^a_{\ b} \Big) \ ,
\end{equation}
where $G_D$ is the bulk (fundamental) Newton's constant. The Weyl
tensor describes gravitational radiation in the bulk, and is
solved for using the bulk Bianchi identities $\nabla_{[a} {\cal
R}_{bc]de}=0$ after expressing the Riemann tensor as a function of
$C_{abcd}$ and ${\bf T}_{ab}$.

After contraction,~(\ref{Gauss equation}) may be expressed as

\begin{eqnarray}
\tilde{G}^m_{\ n} &=& 8\pi G_D\frac{D-3}{D-2}\ \Big[\ {\bf T}^a_{\
b}\ \tilde{g}^m_{\ a} \ \tilde{g}^b_{\ n} -\frac{1}{D-1}\
\tilde{g}^m_{\ n}\ {\bf T}^a_{\ b}\ \tilde{g}^b_{\ a}\Big]+8\pi
G_D \frac{D-3}{D-1}\Big( {\bf T}^a_{\ b}\ n_a\ n^b\ \Big)
\tilde{g}^m_{\ n}
\nonumber \\
&& - E^m_{\ n} + K K^m_{\ n} - K^{m p} K_{pn} - \frac{1}{2}\tilde{g}^m_{\ n}
(K^2 - K_{pq}K^{pq}) \ ,
\label{projected Einstein equations 1}
\end{eqnarray}
where $\tilde{G}_{mn}$ is the Einstein tensor constructed from the induced
metric. In addition to the part linear in the bulk energy-momentum
tensor ${\bf T}_{mn}$, these effective Einstein equations can be
seen to contain source terms quadratic in the brane extrinsic
curvature tensor, plus a tensor contribution, $E_{mn}$, defined from the bulk Weyl tensor
by

\begin{equation}
E_{mn} \equiv C^a_{\ mcn} n_a n^c \ .
\end{equation}
This last term is well known to embody the non-local effects of
the bulk geometry (as opposed to ${\bf T}_{mn}$) on the brane
matter dynamics. Being symmetric and traceless, it can be
interpreted here as ($-8\pi G_D$ times) the energy-momentum tensor
of a conformally invariant source, that we will call from now on
\emph{dark radiation}, borrowing the familiar terminology from the
$D=5$ case applied to cosmological perfect fluids. Loosely
speaking, this may be considered as bulk gravitational radiation.

We now turn to the contributions quadratic in $K_{mn}$. The
extrinsic curvature may be decomposed into its average value and
its jump across the brane (denoted by $[[...]]$) via

\begin{equation}
K_{mn} \ =\ \langle K_{mn}\rangle +\frac{1}{2}\ [[K_{mn}]] \ .
\end{equation}
The jump term is related to the brane-localized energy-momentum
tensor $\tau_{mn}$ by the Israel junction conditions~\cite{junctionconditions}, according to

\begin{equation}
[[K_{mn}]] = -8\pi G_D \Big(\tau_{mn}-\frac{1}{D-2}\ \tau\
\tilde{g}_{mn} \Big) \ ,
\end{equation}
where $\tau \equiv \tilde{g}^{mn}\tau_{mn}$ .

The average value $<K_{mn}>$ on the brane can be extracted by
taking the jump of (\ref{projected Einstein equations 1}), and
using the second of the product rules

\begin{eqnarray}
\langle MN\rangle &=& \langle M\rangle\langle N\rangle + \frac{1}{4}\ [[M]]\ [[N]] \nonumber \\ \ [[MN]]
&=& \langle M\rangle [[N]] + [[M]]\langle N\rangle \ .
\label{product rules}
\end{eqnarray}
Assuming a smooth brane induced geometry, one has
$[[\tilde{G}_{mn}]]=0$, and one obtains \cite{Battye:2001yn}

\begin{eqnarray}
&-\langle K\rangle \tau^m_{\ n}-\frac{1}{D-2}\ \tau \left(\langle K^m_{\ n}\rangle - \langle K\rangle \tilde{g}^m_{\ n}\right) +
\langle K^m_{\ p}\rangle \tau^p_{\ n} + \tau^m_{\ p}\langle K^p_{\ n}\rangle
-\tilde{g}^m_{\ n}\langle K_{pq}\rangle \tau^{pq}
\nonumber \\
&= \left[\left[\frac{1}{8\pi G_D}E^m_{\ n}-\frac{D-3}{D-2}\
\Big({\bf T}^a_{\ b}\ \tilde{g}^m_{\ a} \ \tilde{g}^b_{\
n}-\frac{1}{D-1}\ \tilde{g}^m_{\ n} \ {\bf T}^a_{\ b}\
\tilde{g}^b_{\ a} \Big) - \frac{D-3}{D-1}\ \Big( {\bf T}^a_{\ b}\
n_a\ n^b\ \Big) \ \tilde{g}^m_{\ n}\  \right]\right]
 \nonumber \\
\label{equation for average <Kmn>}
\end{eqnarray}
In~\cite{Battye:2001yn} this was solved for $\langle
K_{mn}\rangle$ in two cases: at linear order in an expansion of
$\tau_{mn}$ around a brane cosmological constant (tension), and
exactly in a cosmological case. However, the latter case involved,
as a formal application, a brane induced geometry with homogeneous
and isotropic (D-2)-dimensional space. In section \ref{section on
cosmological case} we will study a more realistic geometry in
which the brane spacetime is the product of a 4d cosmological one
with a (D-5)-dimensional compact internal space.

The previous literature has mostly focused, in D=5, on the case of
spacetimes which are $\mathbb{Z}_2$-symmetric across the brane, which implies
$\langle K_{mn}\rangle=0$. This has been partly motivated
\cite{Randall:1999ee} by 11-dimensional heterotic M-theory
\cite{Horava:1996ma} with $\mathbb{R}^{10}\times
S^1/\mathbb{Z}_2$ topology. As we will also argue later, this
constraint is less natural for $D>5$, where, for example, an
hyperspherical foliation of the (D-4)-dimensional transverse space
requires $<K_{mn}>\neq 0$ even in the Minkowski bulk vacuum. Note
finally that in certain special cases there is no unique solution
to~(\ref{equation for average <Kmn>}), as can be obviously checked
for a purely geometrical brane with $\tau_{mn}=0$ and no bulk jump on the
right-hand side, for example.

Redefining the energy-momentum tensors to include a bulk
cosmological constant ${\bf \Lambda}$ and a brane tension
$\lambda$,
\begin{eqnarray}
{\bf T}_{ab} &\rightarrow& -{\bf \Lambda}\ g_{ab}+ {\bf T}_{ab}
\nonumber \\
\tau_{mn} &\rightarrow& -\lambda\ \tilde{g}_{mn} + \tau_{mn}\ ,
\label{redefinition to include lambda}
\end{eqnarray}
the Einstein equations on the brane are obtained \cite{Battye:2001yn} by
taking the average value of (\ref{projected Einstein equations 1})
and using the first of the product rules~(\ref{product rules}),
yielding
\begin{equation}
\tilde{G}^m_{\ n}\ =\ \langle \tilde{G}^m_{\ n}\rangle\ = ({\rm bulk})^m_{\ n} + ({\rm
brane\ jump})^m_{\ n} + ({\rm brane\ average})^m_{\ n} - \langle E^m_{\ n}\rangle \ ,
\end{equation}
where

\begin{eqnarray}
({\rm bulk})^m_{\ n} &=& 8\pi G_D\ \frac{D-3}{D-2}\Big[\langle
{\bf T}^a_{\ b}\rangle\ \tilde{g}^m_{\ a} \ \tilde{g}^b_{\
n}-\frac{1}{D-1}\ \tilde{g}^m_{\ n}\ \langle{\bf T}^a_{\
b}\rangle\ \tilde{g}^b_{\ a} \Big]
\nonumber \\
&&+ 8\pi G_D \ \frac{D-3}{D-1} \ \tilde{g}^m_{\ n}\ \langle{\bf
T}^a_{\ b}\rangle\ n_a n^b - 8\pi G_D \frac{D-3}{D-1} \
\tilde{g}^m_{\ n}\ {\bf \Lambda}
\nonumber \\
({\rm brane\ jump})^m_{\ n} &=& \frac{1}{4}(8\pi
G_D)^2\Big[-\frac{D-3}{D-2}\ \frac{\lambda^2}{2}\ \tilde{g}^m_{\
n} + \frac{D-3}{D-2}\ \lambda\ \tau^m_{\ n}
\nonumber \\
&&+\frac{1}{D-2}\tau\tau^m_{\ n}
-\tau^{mp}\tau_{pn}-\frac{1}{2}\tilde{g}^m_{\ n}
\Big(\frac{1}{D-2}\tau^2-\tau^{pq}\tau_{pq}\Big)\Big]
\nonumber \\
({\rm brane\ average})^m_{\ n} &=& \langle K\rangle \langle K^m_{\
n}\rangle - \langle K^{m p}\rangle\langle K_{pn}\rangle -
\frac{1}{2}\tilde{g}^m_{\ n} \left(\langle K\rangle^2 - \langle
K_{pq}\rangle\langle K^{pq}\rangle\right) \
\nonumber \\
&\ & \label{projected Einstein equations 2}
\end{eqnarray}
which was also initially derived in \cite{Shiromizu:1999wj} for
$D=5$ and $\langle K_{mn} \rangle=0$.

As is well known from $D=5$, and remains true here, in the absence of
a brane tension these equations would be quadratic in the brane
energy-momentum $\tau^m_{\ n}$ and thus would be immediately excluded
by observations. However,
a striking result is that for non-zero tension $\lambda$, the
contribution linear in $\tau^m_{\ n}$ has the correct tensor
structure - i.e. no extra terms of the form $\tau\ \tilde{g}^m_{\ n}$.

\subsection{The conservation equations}

Taking the jump and average value across the brane for the Codazzi
equation~(\ref{Codazzi equation}), one obtains
\begin{eqnarray}
\tilde{\nabla}_m \tau^m_{\ n} &=& [[{\bf T}_{ab}]]\ n^a\ \tilde{g}^b_{\ n}
\nonumber \\
\tilde{\nabla}_n \langle K\rangle - \tilde{\nabla}_m \langle K^m_{\ n}\rangle &=& 8\pi G_D \langle{\bf
T}_{ab}\rangle n^a \tilde{g}^b_{\ n} \ .
\end{eqnarray}
The first equation has the obvious meaning that the
non-conservation of the brane energy is given by the jump of the
bulk energy flow across the brane. For realistic applications, we
will assume in the following that there is no such jump and that
$\tau^m_n$ is thus conserved. Note that the Codazzi equation is
essential to obtain this result, as in principle the effective
Einstein equations~(\ref{projected Einstein equations 2}) only
allow one to prove the conservation of its right-hand side as a
whole through the Bianchi identity $\tilde{\nabla}_m
\tilde{G}^m_{\ n} \equiv 0$, not the independent conservation of
every term in the sum.

In a simple configuration with only a cosmological constant ${\bf
\Lambda}$ in the bulk, automatically conserved, and $\langle K_{mn}\rangle=0$,
one obtains from $\tilde{\nabla}_m \tilde{G}^m_{\ n} \equiv 0$ a
(non-)conservation law for the dark radiation:
\begin{equation}
\frac{4}{(8\pi G_D)^2}\tilde{\nabla}_p\ E^p_{\ m} =\frac{1}{D-2}\
\tau^p_{\ m} \tilde{\nabla}_p\ \tau - \tau^p_{\ q}\ \tilde{\nabla}_p
\tau^q_{\ m} - \frac{1}{D-2}\ \tau \tilde{\nabla}_m \tau + \tau^p_{\ q}\
\tilde{\nabla}_m \tau^q_{\ p} \ .
\label{dark radiation conservation law}
\end{equation}
When applying this to the cosmological case, we will see that qualitatively new consequences
appear for $D>5$.

\section{Cosmology with bulk spherical symmetry \label{section on cosmological case}}
\label{bulkspher}
\subsection{The D=5 case}

Although the effective Einstein equations (\ref{projected Einstein
equations 2}) give a very general and covariant result, free of
assumptions about the bulk or brane background, they do not in
general form a closed system, because of the dark radiation source
$E_{mn}$. This last term (and actually the whole bulk Weyl tensor
$C_{abcd}$) needs to be calculated everywhere in the bulk using, in
addition to (\ref{projected Einstein equations 1}),
the Bianchi identities $\nabla_{[a}{\cal R}_{bc]de}=0$ before one
can evaluate it on the brane. This implies solving for the whole
bulk metric. The choice of the bulk topology and the boundary
conditions can strongly affect $E_{mn}$ and, in turn, the evolution of the brane
geometry. It has been shown however
\cite{Binetruy:1999hy}, that for $D=5$, assuming cosmological
symmetries, the $\mathbb{Z}_2$ symmetry and a pure cosmological constant
in the bulk is enough to uniquely determine the dark radiation
term on the brane without having to specify the bulk metric. In this case,
the projective approach described above acquires its full
power. A generalized Birkhoff theorem \cite{Bowcock:2000cq} has
then been established for $D=5$, showing that the bulk geometry
itself is unique with these assumptions, taking the
(A)dS-Schwarschild form, static with a moving brane. This has been
further generalized to non $\mathbb{Z}_2$-symmetric spacetimes
(studied in
\cite{Ida:1999ui,Davis:2000jq,Deruelle:2000ge,Perkins:2000zp,
Stoica:2000ws})
with different bulk cosmological constants and black hole mass
parameters on each side of the brane: the bulk has again to take
the (A)dS-Schwarzchild form on each side. We give here an
alternative proof of the first, brane-based, result regarding the uniqueness
of the dark radiation term for the $\mathbb{Z}_2$-symmetric case.

The contracted Bianchi identities $\nabla_{[a} {\cal R}_{bc]d}^{\
\ \ \ a}$ can be rewritten as
\begin{equation}
\nabla^e C_{abce} = -8\pi G_D\frac{D-3}{D-2} \nabla_{[a} \Big({\bf
T}_{b]c} -\frac{1}{D-1}{\bf T} g_{b]c}\Big) \ .
\label{contracted
Bianchi identities}
\end{equation}
For $D=5$, the dark radiation $E^m_{\ n}$ has only one independent
component, $E^0_{\ 0} = - 3 E^x_{\ x}$ (no sum implied) from 3d isotropy
and tracelessness. The most general bulk metric has the form

\begin{equation}
ds^2_D= -n^2(r,t)dt^2 + a^2(r,t) dx^2_k + b^2(r,t) dr^2 \ ,
\end{equation}
where $dx^2_k$ is the 3-space homogeneous and isotropic line
element with curvature parameter $k$. The brane can be assumed,
without loss of generality, to sit at $r=r_0$. Coordinate freedom
allows one to set $n(r_0,t)\equiv 1$ so that the brane induced
metric has the standard form with cosmic time and scale factor
$a(r_0,t)$. The $r0r$ component of (\ref{contracted Bianchi
identities}) then reads

\begin{equation}
\Big(\partial_0 + 4\frac{\dot{a}}{a}(r,t)\Big)E^0_{\ 0}(r,t) =-8\pi
G_{(5)}. \frac{2}{3}\Big(\nabla_r {\bf T}_0^{\ r} -\nabla_0 ({\bf
T}^r_{\ r}-\frac{1}{D-1}{\bf T})\Big) .
\end{equation}
It is manifest that for ${\bf T}_{ab}=-{\bf \Lambda}g_{ab}$ the
right-hand side vanishes and

\begin{equation}
E^0_{\ 0}(r,t)= \frac{{\cal C}(r)}{a^4(r,t)} \ \ \Rightarrow \ \
E^0_{\ 0}(r_0,t) = -3E^x_{\ x}(r_0,t)=\frac{{\cal C}(r_0)}{a^4(r_0,t)}\ ,
\end{equation}
with ${\cal C}(r)$ an integration constant, so that the dark
radiation is completely specified on the brane.

Alternatively, one may use the conservation law (\ref{dark
radiation conservation law}) and find that the right-hand side is
\emph{identically vanishing} for homogeneous isotropic radiation
in $D=5$, implying conserved radiation and thus $1/a^4$ evolution.
We now turn to the more complex $D>5$ case.

\begin{figure}
\begin{center}
\includegraphics[width=10cm]{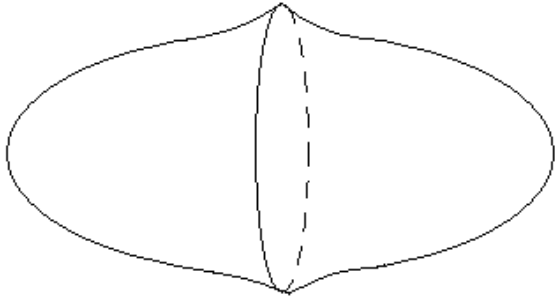}
\end{center}
\caption{Configuration with a $\mathbb{Z}_2$-symmetric compact
bulk, $\langle K_{mn}\rangle=0$, with spherical symmetry around
the brane (pictured for $D=6$). The equatorial circle is the
(D-1)-dimensional brane and the transverse space is compact.}
\label{figure with compact bulk}
\end{figure}

\begin{figure}
\begin{center}
\includegraphics[width=10cm]{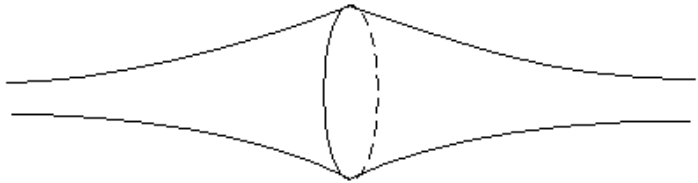}
\end{center}
\caption{Configuration with a $\mathbb{Z}_2$-symmetric non-compact
bulk, $\langle K_{mn}\rangle=0$, with spherical symmetry around
the brane (pictured for $D=6$). The circle is the
(D-1)-dimensional brane and the non-compact space transverse to
the brane has to localize gravity.} \label{figure with non-compact
orbifold}
\end{figure}

\subsection{Arbitrary number of dimensions with $\langle K_{mn}\rangle =0$}

In this case there is considerably more freedom for the bulk
geometry. We consider the most simple and symmetric possibility; a
spherical (or axial) symmetry around the brane. The most general
corresponding bulk metric can be written as
\begin{equation}
ds^2_D=-n^2(r,t)dt^2 + a^2(r,t)dx^2_k + b^2(r,t) dr^2 + h^2(r,t)
r^2 d\Omega^2_{D-5} \ , \label{bulk cosmological+spherical metric}
\end{equation}
where $d\Omega^2$ is the $(D-5)$-dimensional sphere metric. The
brane can again be assumed to sit at $r=r_0$, or have a trajectory
$r(t)$; in any case the $(D-1)$-dimensional brane induced metric
will be

\begin{equation}
\tilde{ds}^2=-d\tilde{t}^2+\tilde{a}^2(\tilde{t})dx^2_k +
R^2(\tilde{t})d\Omega^2_{D-5} \ ,
\end{equation}
where we have used coordinate freedom to use the cosmic time
variable. From now on, we drop the tildes in the metric
coefficients and time coordinate above. This is the product of a
cosmological 4d spacetime with an internal sphere with
time-dependent radius, in which matter is allowed to propagate, as
is familiar from traditional Kaluza-Klein cosmology
\cite{Abbott:1984ba,Abbott:1984nt,Sahdev:1985ye,Sahdev:1988fp}.

We now show, using the same reasoning as in $D=5$, why dark
radiation is no longer uniquely defined from the assumption of 3d
homogeneity and isotropy, and not even after assuming spherical
symmetry around the brane. All the independent contracted Bianchi
identities resulting from (\ref{contracted Bianchi identities})
have been collected in the appendix. In particular, the $r0r$
component reads
\begin{equation}
\nabla_e C_{r0}^{\ \ r e}=\left(\partial_0+3\frac{\dot{a}}{a}\right)C_{r0}^{\
\ r0}+\frac{a'}{a}C_{0k}^{\ \ rk}+\frac{1}{r}C_{0K}^{\ \
rK}-\frac{\dot{a}}{a}C_{rk}^{\ \ rk}=0 \ ,
\end{equation}
when the bulk energy-momentum is covariantly constant, producing
no right-hand side (except a singular distribution on the brane,
omitted here). Using total tracelessness, this becomes
\begin{equation}
\left(\partial_0+4\frac{\dot{a}}{a}\right)E^0_{\ 0}=\left(\frac{a'}{a}-\frac{1}{r}\right)C_{0K}^{\
\ rK}-\frac{\dot{a}}{a}E^K_{\ K} \ ,
\end{equation}
where $K$ spans hypersphere indices. The previously unique $1/a^4$
behavior is manifestly spoiled by the presence of additional
dimensions beyond the fifth, which now require solving for other
components of the Weyl tensor in the bulk.

Taking into account these symmetries, the energy-momentum tensors
can be written as

\begin{eqnarray}
\tau^m_{\ n} &=& {\rm diag}(-\rho,p^x\ \delta^i_{\ j}\ ,p^\theta\
\delta^I_{\ J})
\nonumber \\
-\frac{1}{8\pi G_D} \langle E^m_{\ n}\rangle &=& {\rm diag}(-\rho_d,p^x_d\
\delta^i_{\ j}\ , p^\theta_d \ \delta^I_{\ J})
\nonumber \\
\langle {\bf T}^a_{\ b}\rangle &=& {\rm diag}(-\rho_B,p^x_B\ \delta^i_{\ j}, p^\theta_B
\ \delta^I_{\ J}, p^r_B) + (p^r_0)_B \delta^a_{\ r} \delta_b^{\ 0} +(p^0_r)_B
\delta^a_{\ 0} \delta_b^{\ r} \ ,
\end{eqnarray}
where $i,j=1 .. 3$ denote the 3-space indices and $I,J$ the
hypersphere angle indices. In addition, the dark radiation satisfies

\begin{equation}
\rho_d= 3(p_x)_d + (D-5) (p_\theta)_d \ .
\end{equation}
The dark radiation conservation equation (\ref{dark radiation
conservation law}) for a pure cosmological constant in the bulk is
now

\begin{eqnarray}
\dot{\rho}_d + 3\frac{\dot{a}}{a}\Big(\rho_d&+&(p_x)_d\Big)
+ (D-5)\frac{\dot{R}}{R}\Big(\rho_d+(p_\theta)_d\Big)
\nonumber \\
&=&8\pi
G_D\frac{3}{4}\frac{D-5}{D-2}(p_x-p_\theta)\Big(\dot{p}_x+\frac{\dot{a}}{a}(\rho+p_x)
-\dot{p}_\theta -\frac{\dot{R}}{R}(\rho+p_\theta)\Big) \ .
\label{dark radiation conservation law cosmo}
\end{eqnarray}
This is a new result for $D>5$ brane cosmology: at the
homogeneous background level, dark radiation is no longer
conserved. Brane energy-momentum is conserved in the
higher-dimensional sense, up to a jump term for the bulk energy
flux:

\begin{eqnarray}
\dot{\rho}+3\frac{\dot{a}}{a}(\rho+p_x)
+(D-5)\frac{\dot{R}}{R}(\rho+p_\theta) = -[[(p^r_0)_B]]\ n_r .
\end{eqnarray}

Assuming first that $\langle K_{mn}\rangle=0$ on the brane (see
figures \ref{figure with compact bulk} and \ref{figure with
non-compact orbifold}), the effective Einstein equations are:

\underline{$00$-component}:

\begin{eqnarray}
&&3\left(\frac{\dot{a}}{a}\right)^2 +3 \frac{k_x}{a^2}  +
(D-5)\Big[3\frac{\dot{R}}{R}\frac{\dot{a}}{a}+\frac{D-6}{2}
\left(\frac{\dot{R}}{R}\right)^2+\frac{D-6}{2}\frac{k_\theta}{R^2}\Big]
\nonumber \\
&=& 8\pi G_D \frac{D-3}{D-1}{\bf \Lambda}+\frac{(8\pi G_D)^2}{8}
\frac{D-3}{D-2}\lambda^2
\nonumber \\
&&+ 8\pi G_D \frac{D-3}{D-2}\Big[\rho_B
-\frac{D-2}{D-1}p^r_B+\frac{1}{D-1}(-\rho_B+3
p^x_B+(D-5)p^\theta_B)\Big]
\nonumber \\
&&+8\pi G_D\ \rho_d+\frac{(8\pi G_D)^2}{4}\frac{D-3}{D-2}\
\lambda\ \rho
\nonumber \\
&&+\frac{1}{8}(8\pi
G_D)^2\frac{D-3}{D-2}\Big(\rho^2-3\frac{D-5}{D-3}(p_x-p_\theta)^2\Big)
\ , \label{projected Einstein equations cosmo-I}
\end{eqnarray}

\underline{$xx$-component}:

\begin{eqnarray}
&&2\frac{\ddot{a}}{a}+\left(\frac{\dot{a}}{a}\right)^2
+\frac{k_x}{a^2} +(D-5)\Big[
2\frac{\dot{R}}{R}\frac{\dot{a}}{a}+\frac{D-6}{2}
\left(\frac{\dot{R}}{R}\right)^2+\frac{D-6}{2}\frac{k_\theta}{R^2}+\frac{\ddot{R}}{R}\Big]
\nonumber \\
&=&8\pi G_D \frac{D-3}{D-1}{\bf \Lambda}+\frac{(8\pi G_D)^2}{8}\frac{D-3}{D-2}\lambda^2
\nonumber \\
&&-8\pi G_D \frac{D-3}{D-2}\Big(p^x_B
+\frac{D-2}{D-1}p^r_B-\frac{1}{D-1}[-\rho_B+3
p^x_B+(D-5)p^\theta_B]\Big)
\nonumber \\
&&-8\pi G_D \ p^x_d - \frac{(8\pi G_D)^2}{4}\frac{D-3}{D-2}\
\lambda \ p_x
\nonumber \\
&&-\frac{1}{8}(8\pi G_D)^2\frac{D-3}{D-2}\Big(\rho^2+2\rho\ p_x
+2\frac{D-5}{D-3}(p_\theta-p_x)[\rho+p_x+\frac{3}{2}(p_\theta-p_x)]\Big)
\nonumber \\
&\ &
\label{projected Einstein equations cosmo-II}
\end{eqnarray}

\eject

\underline{$\theta\theta$-component}:

\begin{eqnarray}
&&3\frac{\ddot{a}}{a}
+3\left(\frac{\dot{a}}{a}\right)^2+3 \frac{k_x}{a^2}
+(D-6)\Big[\frac{D-7}{2}\left(\frac{\dot{R}}{R}\right)^2+\frac{D-7}{2}\frac{k_\theta}{R^2}
+3\frac{\dot{a}}{a}\frac{\dot{R}}{R}+\frac{\ddot{R}}{R}\Big]
\nonumber \\
&=&8\pi G_D \frac{D-3}{D-1}{\bf \Lambda}+\frac{(8\pi G_D)^2}{8}\frac{D-3}{D-2}\lambda^2
\nonumber \\
&&-8\pi G_D \frac{D-3}{D-2}\Big(p^\theta_B
+\frac{D-2}{D-1}p^r_B-\frac{1}{D-1}[-\rho_B+3
p^x_B+(D-5)p^\theta_B]\Big)
\nonumber \\
&&-8\pi G_D\ p^\theta_d-\frac{(8\pi G_D)^2}{4}\frac{D-3}{D-2}\
\lambda \ p_\theta
\nonumber \\
&&-\frac{1}{8}(8\pi G_D)^2\frac{D-3}{D-2}\Big(\rho^2+ 2\rho\
p_x+\frac{p_\theta-p_x}{D-3}[2(D-6)\rho-6
p_x+3(D-7)(p_\theta-p_x)]\Big)
\nonumber \\
&\ &
\label{projected Einstein equations cosmo-III}
\end{eqnarray}
Again, $\lambda$ and ${\bf \Lambda}$ are respectively the brane and bulk
cosmological constants defined in (\ref{redefinition to include
lambda}), $k_x$ is the 3-space curvature parameter, and
$k_\theta>0$ similarly for the hypersphere.

The effective Newton constant and cosmological constant on the brane are
thus given by
\begin{eqnarray}
8\pi G_{D-1} &=& \frac{1}{4}\frac{D-3}{D-2}\ \lambda\ (8\pi G_D)^2
\nonumber \\
\Lambda_{D-1} &=& \frac{4}{8\pi G_D} \frac{D-2}{D-1}\frac{{\bf
\Lambda}}{\lambda}+\frac{\lambda}{2} \ .
\label{Lambda_eff et G_eff}
\end{eqnarray}
The cancellation of $\Lambda_{D-1}$ for large brane tension
$\lambda$ of either sign requires a large negative ${\bf
\Lambda}<0$, while positivity of $G_{D-1}$ requires a positive
brane tension $\lambda>0$, as in the $D=5$ case. Note that
recovering the 4d effective Newton and cosmological constants
requires an additional time-dependent factor - the volume ${{\cal
S}(R)}$ of the (D-5)-dimensional sphere:
\begin{eqnarray}
8\pi G_4(t) &=& \frac{8\pi G_{D-1}}{{\cal S}(R)}
\nonumber \\
\Lambda_4(t) &=&\Lambda_{D-1}{\cal S}(R) . \label{Lambda_eff et
G_eff in 4d}
\end{eqnarray}
Although the product $8\pi G_4(t) \Lambda_4(t)=8\pi
G_{D-1}\Lambda_{D-1}$ seems then time-independent, it must be
remembered that we are not in the Einstein frame (E.F.) here, the
4d effective Newton constant being $R-$dependent. One has actually
$\Lambda_4^{E.F.}\sim ({\cal S}(R))^{-1}$, unlike a real
cosmological constant source.

\subsection{Arbitrary number of dimensions with $\langle K_{mn}\rangle \neq 0$}

\begin{figure}
\begin{center}
\includegraphics[width=10cm]{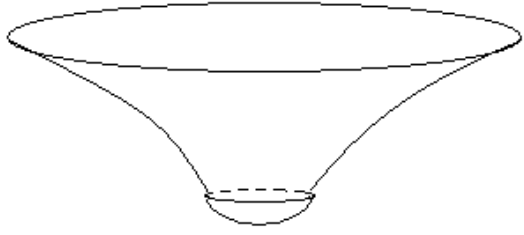}
\end{center}
\caption{Natural configuration with no $\mathbb{Z}_2$ symmetry,
$\langle K_{mn}\rangle \neq 0$, and spherical symmetry around the
brane (pictured for $D=6$). The smaller circle is the
(D-1)-dimensional brane. The bulk space encompassed by the brane
is compact, while the outer space
is non compact and has to localize gravity. The codimension one
brane may be considered as a fat core regularization of a
codimension (D-4) singular brane.} \label{figure with
non-symmetric bulk}
\end{figure}

The assumption of vanishing average extrinsic curvature on the
brane is rather unnatural for $D>5$. It implies configurations
such as those depicted in figure \ref{figure with compact bulk} or
\ref{figure with non-compact orbifold}, but forbids the more
natural configuration of figure \ref{figure with non-symmetric
bulk} where the brane can be considered as located approximately
at the origin of the spherical coordinate system. In this last case
is useful to consider the codimension one brane as a
regularization of a codimension (D-4) singular brane. Furthermore,
a $D-$dimensional Minkowski bulk, described by $n=a=b=h=1$ and
$k=0$ in (\ref{bulk cosmological+spherical metric}), would have a
non-vanishing $\langle K^I_{\ J}\rangle=-\frac{1}{r}\delta^I_{\
J}$ in directions tangent to the hypersphere. Thus, in general we
must include contributions from non-zero $\langle K_{mn}\rangle$
on the right-hand sides of (\ref{projected Einstein equations
cosmo-I}-\ref{projected Einstein equations
cosmo-III}). Such terms are a rational fraction of the brane
energy-momentum, and quadratic in the jump of the bulk
energy-momentum and Weyl tensors. The full expression being rather
complicated, we provide for simplicity only the first terms in a
series expansion in the variable $\rho/\lambda$ ($p_x$ and
$p_\theta$ being of the same order as $\rho$). We assume also a
simple form for the jump

\begin{eqnarray}
&&\left[\left[\frac{1}{8\pi G_D}E^m_{\ n}-\frac{D-3}{D-2}\
\Big({\bf T}^a_{\ b}\ \tilde{g}^m_{\ a} \ \tilde{g}^b_{\
n}-\frac{1}{D-1}\ \tilde{g}^m_{\ n}\ {\bf T}^a_{\ b}\
\tilde{g}^b_{\ a} \Big) - \frac{D-3}{D-1}\ \Big( {\bf T}^a_{\ b}\
n_a\ n^b\ \Big) \ \tilde{g}^m_{\ n}\ \right]\right]
\nonumber \\
&&=\Delta\ \tilde{g}^m_{\ n} \ ,
\end{eqnarray}
which may be realized, for example, if $[[E^m_{\ n}]]=0$ and the bulk
content consists only of cosmological constants with different
values on each side of the brane. The new contributions are then
\begin{eqnarray}
-\tilde{G}^0_{\ 0} &\supset&
\frac{1}{2}\frac{D-2}{(D-3)^2}\frac{\Delta^2}{\lambda^2}\Big[1-2\frac{\rho}{\lambda}
-\frac{3}{\lambda^2}\Big(3(D-5)(p_x-p_\theta)^2-(D-3)\rho^2\Big) +
O(\rho^3/\lambda^3)\Big]
\nonumber \\
-\tilde{G}^x_{\ x} &\supset&
\frac{1}{2}\frac{D-2}{(D-3)^2}\frac{\Delta^2}{\lambda^2}\Big[1+2\frac{p_x}{\lambda}-\frac{3}{\lambda^2}\Big((D-5)(-4p_x
p_\theta+3p_\theta^2+p_x^2+2\rho p_\theta+\rho^2)
\nonumber \\
&&+2\rho p_x+2\rho^2\Big) + O(\rho^3/\lambda^3)\Big]
\nonumber \\
-\tilde{G}^\theta_{\ \theta} &\supset&
\frac{1}{2}\frac{D-2}{(D-3)^2}\frac{\Delta^2}{\lambda^2}\Big[1+2\frac{p_\theta}{\lambda}-\frac{3}{\lambda^2}
\Big(-3(D+1)p_\theta^2+(D-3)\rho^2+3(D-5)p_x^2
\nonumber \\
&&+18p_x p_\theta +6\rho(p_x-p_\theta)\Big) +
O(\rho^3/\lambda^3)\Big] \ .
\label{extra contribution to rhs for
<Kmn> non-zero}
\end{eqnarray}
Note that the contribution linear in $\tau^m_{\ n}$ again has the
correct tensor structure $\tilde{G}^m_{\ n} \supset \tau^m_{\ n}$ with
no $(\tau \tilde{g}^m_{\ n})$ contribution.

\subsection{Evolution of the sources \label{section evolution of sources}}

Following traditional Kaluza-Klein cosmology
\cite{Abbott:1984ba,Abbott:1984nt,Sahdev:1985ye,Sahdev:1988fp},
three simple cases may be considered for the equations of state.
In the \emph{early radiation era} or decompactification regime,
for large temperature $T \gg 1/R$, incoherent radiation is not
sensitive to the compactness of the internal space and is fully
isotropic:

\begin{equation}
p_\theta = p_x= \frac{1}{D-2}\rho \ .
\end{equation}
The brane energy density then evolves according to

\begin{equation}
\rho  \sim \left(\frac{1}{a^3R^{D-5}}\right)^{\frac{D-1}{D-2}} \ ,
\end{equation}
which reproduces a $1/a^4$ behavior for $D=5$, in the absence of an
internal space.

In the \emph{late radiation era}, for $T\ll 1/R$, but before
matter-radiation equality, the temperature is no longer
sufficient to excite an incoherent mixture of Kaluza-Klein modes
on the hypersphere, so that the transverse pressure drops
($p_\theta=0$). Consequently $p_x = \frac{1}{3}\rho$, and the
evolution becomes $\rho \sim 1/(a^4 R^{D-5})$. However, it should
be noted that, due to compactification, the effective energy
density will include a volume factor $\rho_{eff}\sim \rho
R^{D-5}$, so that
\begin{equation}
\rho_{eff}\sim a^{-4} \ ,
\end{equation}
as expected.

Finally, in the \emph{matter era}, after matter-radiation
equality, $p_\theta=p_x=0$, and the brane energy evolves as

\begin{equation}
\rho_{eff}\sim a^{-3} \ .
\end{equation}
Interestingly, this schematic picture may also be applied to
dark radiation. In the early radiation and matter eras one has
$(p_x)_d=(p_\theta)_d$ which, according to (\ref{dark
radiation conservation law cosmo}), leads to independent conservation.
The evolution is the same as that of brane energy. In the late
radiation era, $p_x \neq p_\theta$ and the non-conservation
equation reads

\begin{equation}
\partial_t \left(\ \rho^{eff}_d \ a^{3\frac{D-1}{D-2}}\ R^{\frac{D-5}{D-2}}\
\right)= -\frac{D-5}{4}\ 8\pi G_D\
a^{3\frac{D-1}{D-2}}R^{\frac{D-5}{D-2}}\Big[(\rho^{eff}_0)^2
\left(\frac{a_0}{a}\right)^8\Big]\frac{1}{{\cal S}(R)}\frac{\dot{R}}{R} \ ,
\end{equation}
where ${\cal S}(R)\sim R^{D-5}$ is the area of the
(D-5)-dimensional sphere, and $\rho_0^{eff}$, $a_0$ are the values
of the effective brane energy density (not of the dark radiation)
and scale factor at an arbitrary reference time. This equation is
not sufficient to solve for $\rho^{eff}_d (a,R)$, unless one assumes an
algebraic relation between the two scale factors $a$
and $R$.

\subsection{An example with a non-compact orbifold bulk}

We have not yet made reference to a specific bulk geometry. As we
mentioned before, the effective Einstein equations given
previously do not guarantee the existence of a bulk solution. In
the non-compact case in particular, the bulk must lead to localized
gravity: in the absence of brane matter there must exist a finite volume
static solution.
Unfortunately, it has been shown\footnote{At least for the case of
a decreasing exponential 4d scale factor $ds^2 \supset
\exp(-\alpha r)\eta_{\mu\nu}dx^\mu dx^\nu +dr^2$.} in
\cite{Gherghetta:2000jf} that for a pure cosmological constant in
the bulk, a compact space encompassed by the brane and a finite
support for the brane energy-momentum, no localization occurs in
$D>6$. This includes the configuration of figure \ref{figure with
non-symmetric bulk}, but note that the case of figure \ref{figure
with non-compact orbifold} is not covered by this theorem.
Therefore, for the figure \ref{figure with non-symmetric bulk}
case, a non-trivial bulk energy-momentum content is required,
which has to extend to infinity, in addition to the infinitely
thin brane located at finite $r$. In the following example, even
though we assume a $\mathbb{Z}_2$ symmetry so that localized gravity
may still be possible with a trivial bulk source, we
will nevertheless use a non-cosmological constant bulk source.

A simple possibility \cite{Olasagasti:2000gx} is a global
topological defect made of a set of $D-5$ scalar fields $\phi^a$
with a potential
\begin{equation}
V(\phi)\sim\left(\sum_a \phi^a \phi^a-v^2\right)^2 \ ,
\end{equation}
minimized in the ``hedgehog" configuration $\phi^a = v\
u^a(\theta^K)$, with $v$ a constant scale and $u^a$ the unit
radial vector field.
The associated hyperspherical metric is then
\begin{equation}
d\Omega^2_{D-5}=\sum_{I=1}^{D-5} \Big(\prod_{K=1}^{I-1}
\sin^2\theta_K\Big) d\theta_I^2
\end{equation}
and the corresponding energy-momentum source, isotropic in both the
3-space and in the hypersphere, is
\begin{eqnarray}
8\pi G_D{\bf T}^\mu_{\ \nu}&=& -\frac{D-5}{D-2}\ \frac{8\pi
G_Dv^2}{8\pi G_D v^2-(D-6)}\ {\bf\Lambda}\ \delta^\mu_{\ \nu}
\nonumber \\
8\pi G_D{\bf T}^r_{\ r}&=&-\frac{D-5}{D-2}\ \frac{8\pi G_Dv^2}{8\pi
G_D v^2-(D-6)}\ {\bf\Lambda}
\nonumber \\
8\pi G_D{\bf T}^I_{\ J} &=&  -\frac{D-7}{D-2}\ \frac{8\pi G_Dv^2}{8\pi
G_D v^2-(D-6)}\ {\bf\Lambda}\ \delta^I_J \label{hedgehog source}
\end{eqnarray}
in addition to the bulk cosmological constant ${\bf\Lambda}<0$. (Recall
that $\mu,\nu=0..3$ and $I,J$ are hypersphere angle indices.)

For $8\pi G_D v^2>(D-6)$, the static background metric is

\begin{equation}
ds^2_D=\exp\Big(-\sqrt{\frac{-16\pi G_D{\bf\Lambda}}{D-2}}
r\Big)(-dt^2+\delta_{ij}dx^i dx^j)+dr^2 + \frac{D-2}{-16\pi
G_D{\bf\Lambda}}\Big(8\pi G_D v^2-(D-6)\Big)d\Omega^2_{D-5} \ .
\end{equation}
As the metric coefficient of the sphere line element
$d\Omega_{D-5}^2$ is constant and thus does not vanish at $r=0$,
the topology of the (D-4)-dimensional space transverse to the 4d
visible dimensions is that of a cylinder $\mathbb{R}\times
S^{D-5}$ rather than $\mathbb{R}^{D-5}$. The defect is located at
$r=+\infty$ in these coordinates. Forgetting the topological
defect origin of this bulk source, it may be considered as the
most simple generalization of a bulk cosmological constant,
(dropping the requirement of D-dimensional local Lorentz invariance) obeying
4d local Lorentz invariance with (D-4)-dimensional
local isotropy.

If we let $r$ extend to $-\infty$ in the metric above, localization of gravity
is lost. If the brane, in the static case, is located at $r=0$,
we need a warp factor decreasing exponentially as
$r\rightarrow -\infty$. As is familiar from $D=5$, an adequate
partially anisotropic background brane tension produces the wanted
effect. To simplify, we may assume a $\mathbb{Z}_2$ symmetry across the
brane; simply replacing $r$ with $|r|$ in the metric
above. The corresponding brane tension
$(\lambda_x,\lambda_\theta)$ is then obtained from the junction conditions:

\begin{eqnarray}
\rho &=&-p_x = 3\sqrt{\frac{-16\pi G_D{\bf\Lambda}}{D-2}} \equiv
\lambda_x
\nonumber \\
p_\theta &=& -4\sqrt{\frac{-16\pi G_D{\bf\Lambda}}{D-2}}\equiv
-\lambda_\theta \ .
\end{eqnarray}
Alternatively, this may be deduced from the point of view of the
effective Einstein equations on the brane (\ref{projected Einstein
equations cosmo-I}-\ref{projected Einstein
equations cosmo-III}), by taking ${\bf \Lambda}$ plus (\ref{hedgehog
source}) as the bulk source, requiring staticity
($\dot{a}=\dot{R}=0$) and flat 3-space sections ($k_x=0$), setting the
$S^{D-5}$ radius $R$ to
\begin{equation}
\frac{k_\theta}{R^2}=\left(\frac{1}{D-2}\right)\frac{-16\pi
G_D{\bf\Lambda}}{8\pi G_D v^2-(D-6)} \ ,
\end{equation}
and solving for the corresponding brane source.

If we now temporarily restore the metric form $|r|\rightarrow r$
and let the brane move in this bulk with an equation $r(t)$, while
keeping the $\mathbb{Z}_2$ symmetry across the brane, we obtain a
cosmological solution, with induced metric
\begin{equation}
\tilde{ds}^2_{D-1} = -dT^2+a^2(T)\delta_{ij}dx^i dx^j
+\frac{D-2}{-16\pi G_D{\bf\Lambda}}\Big(8\pi G_D v^2-(D-6)\Big)
d\Omega_{D-5}^2 \ .
\end{equation}
Note that the radius of the internal space is not dynamical
($\dot{R}=0$), even when the brane moves in the bulk. After
redefining the brane energy-momentum as a departure from the
tension $\lambda_{x,\theta}$ :
\begin{eqnarray}
\tau^0_{\ 0} &\rightarrow& -\lambda_x -\rho
\nonumber \\
\tau^i_{\ j} &\rightarrow& (-\lambda_x +p_x)\delta^i_{\ j}
\nonumber \\
\tau^I_{\ J} &\rightarrow& (-\lambda_\theta + p_\theta)\delta^I_{\ J}
\end{eqnarray}
one obtains from (\ref{projected Einstein equations cosmo-I}-\ref{projected Einstein equations cosmo-III}):
\begin{eqnarray} 3\left(\frac{\dot{a}}{a}\right)^2 \supset \frac{(8\pi
G_D)^2}{4(D-2)}\Big[(D-3)\lambda_x \rho + 3
(D-5)(\lambda_x-\lambda_\theta)(p_x-p_\theta)\Big] +
O\left(\frac{\rho^2}{\lambda_{x,\theta}^2}\right)
\end{eqnarray}
for the part linear in the brane energy-momentum. The appearance
of the pressures $p_x$ and $p_\theta$ in the Friedmann equation renders
this specific configuration unrealistic;
more generally, the brane effective Einstein equations do not
possess the correct tensor structure. The problem is due to our
expanding the brane energy-momentum around a brane tension
$(\lambda_x,\lambda_\theta)$ that does not enjoy the full
(D-2)-dimensional isotropy. Requiring this is a strong constraint
on the possible configurations leading to realistic gravity. In
the next section, we consider a case with fully isotropic brane
tension, at the expense of dropping spherical symmetry around
the brane.

Another difficulty with the previous example is that the equation
of state is overconstrained:
\begin{equation}
p_\theta =f(\rho,p_x) .
\end{equation}
This constraint is unphysical as soon as an explicit model for the
source is given, in the spirit of section \ref{section evolution
of sources}. Furthermore, this problem is common to all
configurations in which the brane source is described by three or
more parameters (like $\rho$, $p_x$, $p_\theta$ here) while the
brane induced metric has a single independent scale factor $a(t)$,
and this is thus generic in models where a codimension one brane
moves in a static bulk. The conclusion then, is that \emph{a setup
with no overconstraint on the brane equations of state requires a
non-static bulk}. We will comment further on this problem is the
next section.

\section{Cosmology with an AdS-Schwarzschild bulk}
\label{adsbulk}

\subsection{The Flat foliation}

In the previous section we assumed spherical symmetry.
Although this is the most symmetric bulk geometry, it introduces
the complication of having to handle a curved internal space on
the brane - the (D-5)-dimensional hypersphere - which leads to
difficulties in obtaining localized gravity without a
non-trivial bulk source. In addition to this extra complexity,
this led in the previous example to an anisotropic brane
tension, implying an unrealistic effective gravity due to the
wrong tensor structure of the energy-momentum on the right-hand
side of the Einstein equations.

However, replacing this curved space by a flat one, such as
a (D-5)-dimensional hypertorus, allows for a very simple
$AdS_D$-Schwarzschild solution:
\begin{eqnarray}
ds^2_D &=&
-\Big(\frac{-2{\bf\Lambda}}{(D-1)(D-2)}r^2-\frac{\mu}{r^{D-3}}\Big)dt^2+
\frac{dr^2}{\frac{-2{\bf\Lambda}}{(D-1)(D-2)}r^2-\frac{\mu}{r^{D-3}}}
\nonumber \\
&&+r^2 (\delta_{ij}dx^i dx^j+ \delta_{IJ}dX^I dX^J) \ ,
\label{AdS-Schwarzschild with flat sections}
\end{eqnarray}
with a pure cosmological constant source ${\bf\Lambda}<0$ in the
bulk, and where the $X^I$ coordinates are toroidally compactified,
$X^I\in [0,2\pi R_I]$. Time-dependence of the induced metric is
generated by the brane motion $r(t)$ in this static bulk. A
background brane tension
\begin{equation}
\lambda=\sqrt{-8\frac{D-2}{D-1}\frac{{\bf\Lambda}}{8\pi G_D}}
\end{equation}
is then required by the junction conditions to cancel
$\Lambda_{eff}$, in accordance with (\ref{Lambda_eff et G_eff}).
This configuration automatically inherits localized gravity
from the known $AdS_5$ case \cite{Randall:1999vf,Bao:2005ni}.

The
brane induced metric has the form
\begin{equation}
\tilde{ds}^2_{D-1} = -dT^2 + a^2(T) \Big(\delta_{ij}dx^i dx^j +
\sum_{I=1}^{D-5} R_I^2 d\theta_I^2\Big) \ ,
\label{brane metric for
AdS-Schw. bulk with flat sections}
\end{equation}
with $\theta_I \in [0,2\pi]$. In the simplifying case of an
homogeneous hypercube ($R_I=R_1 \ \forall I$), the corresponding
effective Einstein equations (\ref{projected Einstein equations
cosmo-I}-\ref{projected Einstein equations
cosmo-III}) derived in the hyperspherical case still apply, with
$k_x=k_\theta=0$, vanishing bulk source except ${\bf \Lambda}$,
and $R(t)=R_1\ a(t)$. $p_\theta$ is then the brane pressure in the
hypercube directions. This is immediately generalized to
independent radii $R_I$ for the hypertorus, by allowing also for
independent $p_{\theta_I}$.

Dark radiation arises from the bulk Weyl curvature due to the bulk
``black hole mass" parameter $\mu$. In the coordinate system above
\cite{Battye:2001yn}:
\begin{equation}
E^m_{\ n} =-\frac{D-3}{2}\frac{\mu}{a^{D-1}(T)}
\Big(\delta^m_{\ n}-(D-1)\delta^m_{\ 0}\delta_n^{\ 0}\Big)
\end{equation}
for $m,n = 0,x^i,\theta_I$. For differing black hole mass
parameters $\mu$ and/or cosmological constants ${\bf \Lambda}$ on
each side of the brane, the extra contribution (\ref{extra
contribution to rhs for <Kmn> non-zero}) would also appear in the
effective Einstein equations.

Even in absence of dark radiation ($\mu=0$), this configuration
is not viable at late times. Interestingly, this is true
no matter how small the compactification radii $R_I$ are taken, as
long as the rate of variation of the size of the compact space is the same
as that of the macroscopic 3-space,
$\frac{\dot{R}}{R}=\frac{\dot{a}}{a}$, which is implied by the
embedding we have considered.
This may be seen from the brane energy conservation
equation for $\rho_{eff}= \Big(R_1 a(t)\Big)^{D-5}\rho$:
\begin{equation}
\dot{\rho}_{eff} + 3\frac{\dot{a}}{a}(\rho_{eff}+p_x^{eff}) =
-(D-5)\frac{\dot{a}}{a} p_\theta^{eff} \ .
\label{energy conservation
on the brane cosmo case}
\end{equation}
Due to the (D-1)-dimensional local isotropy of the bulk, preserved
by the brane embedding, the junctions conditions imply
$p_x=p_\theta$ on the brane, as in the case considered formally in
\cite{Battye:2001yn} where no realistic compactification was
attempted. In addition to being an unphysical overconstraint on
the equation of state, this leads to an unacceptably large
non-conservation of energy in (\ref{energy conservation on the
brane cosmo case}) compared to standard cosmology, or stated
differently, an unacceptably anomalous evolution law
$\rho_{eff}(t)$. This immediate observational problem may have
been avoided in a bulk configuration in which a small enough
$p_\theta \ll \rho$ is allowed. Again, this problem of a
constraint on $p_\theta$ could be predicted from the existence of
a bulk timelike Killing vector.

In any case, another immediate problem, even if a vanishing
$p_\theta$ was possible, is the unacceptably high rate of variation of the
4d effective Newton's constant, given here in order of magnitude
by the expansion rate:
\begin{equation}
\left|\frac{\dot{G_4}}{G_4}\right| =
\left|-(D-5)\frac{\dot{R}}{R}\right| =
\left|-(D-5)\frac{\dot{a}}{a}\right| \sim 10^{-10}\ {\rm yr}^{-1}
\gg O(10^{-12})\ {\rm yr}^{-1} \ ,
\end{equation}
where we have imposed the present upper bound in particular from
lunar laser ranging \cite{Will:2005va,Williams:2004qb}. This is a
universal constraint independent of the specifics of the
non-gravitational interactions. Of course, if we assume these to
be the Standard Model interactions, there are even stronger
constraints on the cosmological variation of the internal radius,
arising in particular from bounds on the variation of the fine
structure constant. In addition, at the homogeneous level this
study does not capture the effects of the fluctuations of the
unstabilized radion.

\subsection{The Hyperspherical foliation}

We may choose a different parametrization of the
AdS-Schwarzschild bulk (\ref{AdS-Schwarzschild with flat
sections}) using (D-2)-dimensional spheres:
\begin{eqnarray}
ds^2_D &=&
-\Big(1+\frac{-2{\bf\Lambda}}{(D-1)(D-2)}r^2-\frac{\mu}{r^{D-3}}\Big)dt^2+
\frac{dr^2}{1+\frac{-2{\bf\Lambda}}{(D-1)(D-2)}r^2-\frac{\mu}{r^{D-3}}}
\nonumber \\
&&+r^2
\Big[\sum_{I=1}^{D-5}(\prod_{K=1}^{I-1}\sin^2\theta_K)d\theta_I^2
+(\prod_{L=1}^{D-5}\sin^2\theta_L)d\Omega_3^2\Big] \ ,
\end{eqnarray}
where this time $d\Omega_3^2$ is the line element of a 3-sphere
representing the visible macroscopic 3-space. The brane embedding
is now physically different; in particular there is now no need to
compactify the transverse space induced on the brane because the
total brane space is compact. The visible 3-space is now compact
too. The induced metric has the form
\begin{equation}
\tilde{ds}^2_{D-1} = -dT^2 + a^2(T)
\Big[\sum_{I=1}^{D-5}\left(\prod_{K=1}^{I-1}\sin^2\theta_K\right)d\theta_I^2
+\left(\prod_{L=1}^{D-5}\sin^2\theta_L\right)d\Omega_3^2\Big] \ .
\end{equation}
It is obvious that, contrary to the previous case with a flat
foliation, there is no freedom to make the characteristic size of the
internal space much smaller than the radius of the visible 3-space. This immediately
leads to a problem, since we require matter to propagate in
the internal space on the brane.

\section{Conclusions}
\label{conclusions}

In this paper, we have derived the effective Einstein equations
for the cosmological evolution of a codimension one brane embedded
in a general D-dimensional bulk with spherical symmetry around the
brane. The geometrical side of the effective Friedmann equation
corresponds to the product of a 4d cosmological spacetime with an
internal sphere with time-dependent radius. When the brane tension
$\lambda$ enjoys a full (D-2)-dimensional isotropy, the source
side of the Friedmann equation contains the standard term linear
in the brane energy plus quadratic corrections of order
$O(\rho^2/\lambda^2)$, in addition to terms linear in the bulk
energy-momentum. When it does not, the effective gravity does not
have the correct tensor structure for the part linear in the brane
source. We have included contributions to the effective source
arising from the breaking of the $\mathbb{Z}_2$ symmetry across
the brane. We have shown that the bulk radiation manifests itself
as an extra conformally invariant ``dark radiation" source, which
is in general not conserved, contrary to the $D=5$ case. By taking
the example of an AdS-Schwarzschild bulk, we have illustrated the
problem that any static bulk necessarily leads to an
overconstraint on the brane matter equation of state.

These simple analytic examples provide a useful
guide for the explicit construction of realistic cosmological metrics
in the bulk in $D\geq 7$. As an alternative approach to
codimension one branes, it may be interesting to derive the
effective Friedmann equations in a regularization-independent way
for fat branes of codimension (D-4), as a perturbation expansion
in the brane energy density integrated over the core.
\\
\
\\

{\large \bf Acknowledgments}
\\

NC and MT are supported by the National Science Foundation under grant number PHY-0354990, by
Research Corporation, and by funds provided by Syracuse University. NC and CM are supported by the Department of Energy, under grant number DE-FG02-85ER40231.

\section*{Appendix : Bulk contracted Bianchi identities
in the cosmological case\label{appendix on cosmo Bianchi
identities}}

The most general metric with cosmological symmetries and spherical
symmetry around the brane (for vanishing 3-space curvature) can be
written as
\begin{equation}
ds^2_D=-n^2(r,t)dt^2+a^2(r,t)\delta_{ij}dx^i dx^j
+b^2(r,t)dr^2+r^2 d\Omega^2_{D-5} \ ,
\end{equation}
where it is always possible to set the $g_{IJ}$ component,
proportional here to $r^2$, as time-independent, as long as the
brane is allowed to move in the bulk. The contracted Bianchi
identities read, after using Einstein equations,
\begin{equation}
\nabla^e C_{abce} = -8\pi G_D\frac{D-3}{D-2} \nabla_{[a} \Big({\bf
T}_{b]c} -\frac{1}{D-1}{\bf T} g_{b]c}\Big) \ .
\end{equation}
Symmetries severely constrain the 3-index tensor on the
left-hand side, leaving only six independent components (before
using the extra tracelessness constraints). We write them explicitly here:

\begin{eqnarray}
\nabla_e C_{r0}^{\ \ r
e}&=&\left(\partial_0+3\frac{\dot{a}}{a}\right)C_{r0}^{\ \
r0}+\frac{a'}{a}C_{0k}^{\ \ rk}+\frac{1}{r}C_{0K}^{\ \
rK}-\frac{\dot{a}}{a}C_{rk}^{\ \ rk}
\nonumber \\
\nabla_e C_{0r}^{\ \
0e}&=&\left(\partial_r+3\frac{a'}{a}+\frac{D-5}{r}\right)C_{r0}^{\ \
r0}+\frac{\dot{a}}{a}C_{rk}^{\ \ 0k}-\frac{a'}{a}C_{0k}^{\ \
0k}-\frac{1}{r}C_{0I}^{\ \ 0I}
\nonumber \\
\nabla_e C_{i0}^{\ \ je}&=&\left(\partial_0
+\frac{\dot{b}}{b}+2\frac{\dot{a}}{a}\right)C_{i0}^{\ \ j0}+\left(\partial_r
+\frac{b'}{b}+2\frac{a'}{a}-\frac{n'}{n}+\frac{D-5}{r}\right)C_{i0}^{\ \
 jr}-\frac{\dot{a}}{a} C_{ik}^{\ \ jk}-\frac{\dot{b}}{b}C_{ir}^{\ \ jr}
\nonumber \\
\nabla_e C_{I0}^{\ \
Je}&=&\left(\partial_0+\frac{\dot{b}}{b}+3\frac{\dot{a}}{a}\right)C_{I0}^{\ \
J0}+\left(\partial_r+\frac{b'}{b}+3\frac{a'}{a}-\frac{n'}{n}+\frac{D-6}{r}\right)C_{I0}^{\
\ Jr}-\frac{\dot{a}}{a}C_{Ik}^{\ \ Jk}-\frac{\dot{b}}{b}C_{Ir}^{\
\ Jr}
\nonumber \\
\nabla_e C_{ir}^{\ \
je}&=&\left(\partial_0+2\frac{\dot{a}}{a}-\frac{\dot{b}}{b}+\frac{\dot{n}}{n}\right)C_{ir}^{\
\ j0}+
\left(\partial_r+2\frac{a'}{a}+\frac{n'}{n}+\frac{D-5}{r}\right)C_{ir}^{\ \
jr}-\frac{a'}{a}C_{ik}^{\ \ jk}-\frac{1}{r}C_{iK}^{\ \ jK}
\nonumber \\
&&-\frac{n'}{n}C_{i0}^{\ \ j0}
\nonumber \\
\nabla_e C_{Ir}^{\ \
Je}&=&\left(\partial_0+3\frac{\dot{a}}{a}+\frac{\dot{n}}{n}-\frac{\dot{b}}{b}\right)C_{Ir}^{\
\
J0}+\left(\partial_r+3\frac{a'}{a}+\frac{n'}{n}+\frac{D-5}{r}\right)C_{Ir}^{\
\ Jr}-\frac{1}{r}C_{IK}^{\ \ JK}-\frac{n'}{n}C_{I0}^{\ \
J0}
\nonumber \\
&&-\frac{a'}{a}C_{Ik}^{\ \ Jk}
\end{eqnarray}
where $i,j,k$ denote 3-space indices, and $I,J,K$ hypersphere
angle indices.

\end{document}